\documentclass[times,11pt]{sctc10}
\usepackage[latin1]{inputenc}
\usepackage[english]{babel}
\usepackage{times}
\usepackage{amsmath}
\usepackage{amssymb}
\usepackage{graphicx}
\usepackage{subfigure}
\usepackage{moreverb}
\usepackage{geometry}

\usepackage{graphicx}
\usepackage{times}

\renewcommand{\vec}[1]{\boldsymbol{#1}}
\newcommand{\ind}[1]{_{\rm #1}}
\title{SPACECRAFT WAKES IN THE SOLAR WIND}

\author{Anders~I.~ERIKSSON~(1), Yuri~KHOTYAINTSEV~(1) and 
Per-Arne~LINDQVIST~(2)}
\affiliation{(1) Swedish Institute of Space Physics, Uppsala, Sweden \\
(2) Space and Plasma Physics, School of Electrical Engineering, Royal Institute of Technology,
  Stockholm, Sweden
}

\abstract{The solar wind flow creates a wake behind any spacecraft 
immersed in it. We study the properties of this wake using the 
spherical electrostatic probes of the Electric Fields and Waves 
(EFW) instruments on the Cluster satellites. The satellites spin in 
a plane inclined only a few degrees with respect to the ecliptic plane. 
The solar wind is often close to this plane, so each probe 
(44 m away from the spin axis) passes through the wake once 
every spin period (around 4 s), thereby sampling a cut of the 
wake electrostatic potential structure. The signature of the wake 
is clearly seen in the data as a pulse with an amplitude 
typically of a few tenths of a volt. We present statistics of 
the wake signatures as well as detailed examples, compare to 
solar wind parameters, and show a method to remove the wake 
signature from the electric field measurements.}

\begin{document}
\maketitle

\section{INTRODUCTION}

A wake necessarily forms behind any object in a supersonic flow of
plasma or neutral gas. In space plasmas, spacecraft usually
encounter mesosonic flows, i.e.\ flows which are supersonic
with respect to the ion thermal speed, but subsonic with respect
to the electron thermal speed. The result is that the wake charges
negatively until the potential in the wake is sufficiently negative
to prohibit further accumulation of electrons. 

A substantial litterature on spacecraft wake formation has accumulated
over the years. The basic theoretical understanding was 
summarized already by Al'pert et al.\ \cite{Alpert1965a} and
Gurevich et al.\ \cite{Gurevich1969a} in the sixties, though 
significant extensions have also been achieved in later times,
particularly regarding laboratory work,
on-orbit investigations and numerical
simulations. Most work have concentrated on wakes behind
objects at negative potentials with respect to the plasma,
as is relevant for satellites in the ionosphere. There are
several reasons for this: wakes are most conspicuous in 
ionospheric data, where all spacecraft move in the mesosonic
regime almost always; in the ionosphere, spacecraft are larger
than the Debye length, giving large wakes charged to potentials
of order $KT\ind{e}/e$;
wakes in the ionopsphere have influence on the
charging of spacecraft and thus are of interest for spacecraft
design and operations; 
and most satellites, particularly the
bulk of all manned spacecraft, operate in the ionosphere. 

Though it was realized early \cite{Whipple1965a} that spacecraft
in the solar wind would generally be positive, wake formation around
positive spacecraft potentials has therefore recieved little
attention until recently. There are a few exceptions,
including early linearized analytical models \cite{Kraus1958a,Rand1960a}
of uncertain validity \cite[p.~91]{Alpert1965a}, 
laboratory investigations \cite{Hester1970a} and some numerical work, old
\cite{Isensee1977a,Isensee1981a} as well as more recent 
\cite{Singh1994c,Singh1997a,Rylina2002a,Zinin2004a,Roussel2004a}. 
Nevertheless, for satellites doing scientific measurements in
less dense regions, like the magnetosphere above an altitude of
a few thousand kilometers and all of the solar wind, the case
of wakes around positive spacecrafts is important.
Mesosonic flows occur in at least two regions above the ionosphere.
The polar wind is a mesosonic outflow from the Earth into the 
lobes of the magnetotail.
Because the polar wind ion flow energy $m u^2/2$ (which may be below or
around 10~eV) is not sufficient
to overcome the spacecraft potential $V\ind{S}$, 
which can be several tens of volts
positive because of the low density in the lobes \cite{Pedersen1995a},
the ions scatter on the potential structure around the spacecraft,
so that wakes forming behind spacecraft in the polar wind 
can have dimensions much larger than the spacecraft size. The
exploration of this topic is very recent \cite{Eriksson2006a,Engwall2006b},
as reported elsewhere in the present volume \cite{Eriksson2007a}.

The other region above the ionosphere were mesosonic flows are
commonly encountered is the solar wind.
While a satellite in the solar wind develops a
positive potential $V\ind{S}$of up to or around 10~V due to the low 
plasma density, the ion
flow kinetic energy $m u^2/2$ of around 1~keV will always be sufficient
for letting the ions reach the spacecraft. The wake forming
in such a situation will thus be narrow, but it will, as we will
see, still be detectable.
We use the term {\it narrow wake} for this wake, forming
when $m u^2/2 > e V\ind{S}$, to discriminate from the much
broader {\it extended wake} discussed in \cite{Eriksson2007a} that
forms when $m u^2/2 < e V\ind{S}$. 

\begin{figure}
\centerline{
	 \includegraphics[angle=-90,width=8.0cm]{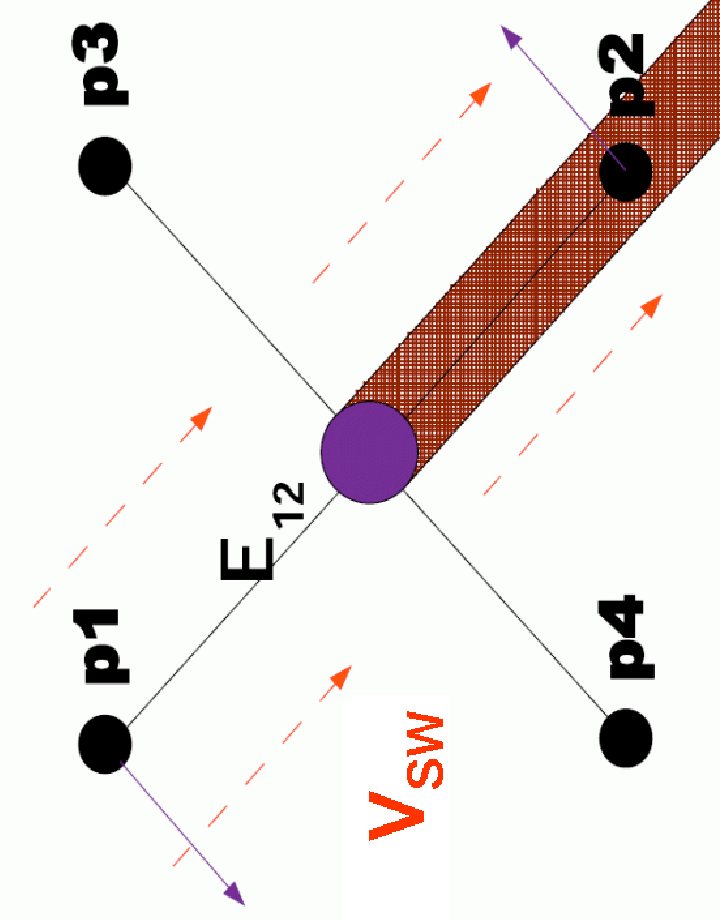}
}
\caption{Sketch of Cluster and EFW probe geometry, with the wake
  forming behind the spacecraft. In reality, the wake will spread
  out at an angle related to the flow Mach number.}
\label{fig:sketch}
\end{figure}

We base this study on data from the Electric Fields and
Waves (EFW) instruments on ESA's four Cluster satellites. Each
EFW instrument uses four spherical sensors at the tip of wire booms
arranged as an orthogonal cross in the satellite spin plane. 
The distance between opposing
probes is 88~m, and the spin period is close to 4~s. 
Further
instrument information is provided in references \cite{Gustafsson1997a}
and \cite{Pedersen1998a}, the latter also giving a thorough
introduction to the measurement technique. 
Figure~\ref{fig:sketch}
shows a sketch of wake generation behind Cluster and its detection
by EFW.
The Cluster satellites
\cite{Escoubet2001a} were launched by ESA in the 
summer 2000 into a polar orbit of
roughly $4\times20~R\ind{E}$ and have by now acquired
detailed four-point measurments in 
the terrestrial magnetosphere and the adjacent
solar wind for the declining half of one solar cycle. 
Apart from EFW data, we will also use
supporting data from the Cluster Ion Spectrometers (CIS, \cite{Reme2001ab}),
particulary the Hot Ion Analyzer (HIA) 
instruments in this paper.

\section{EXAMPLE DATA}\label{sec:ex}

\begin{figure}[t]
\centerline{
	 \includegraphics[angle=-90,width=11.0cm]{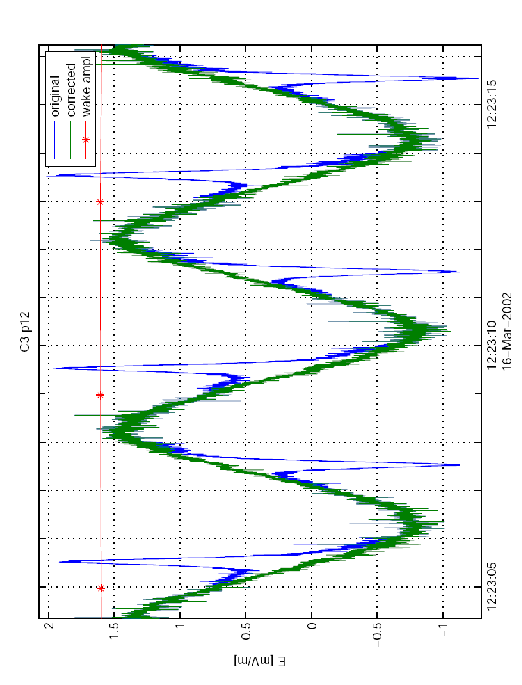}
}
\caption{Example of solar wind wake signature in Cluster EFW data.
  The blue curve is the original raw data, while the green curve
  shows the data after wake removal (Section~\ref{sec:fix}). The
  red stars, bound together by the red curve, shows the wake
  amplitude determined in the removal process, once for each spacecraft
  spin period.} 
\label{fig:ex}
\end{figure}

Figure \ref{fig:ex} shows an example of a narrow wake, acquired by
the EFW instrument on Cluster~3 (that is, number~3 of the four Cluster
spacecraft) in the solar wind.
The data is the electric field measurement (probe voltage difference)
from two opposing EFW probes during 12~s, or three spacecraft spin
periods, in the spinning spacecraft frame of reference. As expected,
the effect of the spacecraft spin is obvious in this rotating frame
of reference, modulating the electric field with a
sinusoid of period close to 4~s. The blue curve shows the original
data. Twice every spin, a narrow ($< 0.5$~s) pulse can be clearly
seen in the data, positive once and negative once per spin. The
pulse pattern is very repetitive, and positive and negative
pulses are approximate mirror images of each other.

The artificial nature of these pulses is very clear, not least from
the perfect repetition at spacecraft spin frequency. 
At first glance, the pulses may be thought to those seen in
electric field data from the EXOS-D/Akebono satellite
by Amano et al.\ \cite{Amano1998a}.
However, we will show that any similarities are incidental, 
and the actual mechanisms
behind the pulse-like signatures in Cluster and Akebono electric field
data are very different. In the case of Akebono, which is always inside
the magnetosphere with an apogee of 10~500~km, the reason presented
by Amano et al.\ is that photoelectrons from the spacecraft, whose
potential is a few volts positive, reach the probes when the wire 
booms are aligned to the ambient magnetic field. The Akebono
electric field results
are consistent with the previous investigations of Langmuir
probe data from the Viking satellite by Hilgers et al.\ 
\cite{Hilgers1992a,Hilgers1993a,Hilgers1995a} 
in an orbit similar to Akebono's, which were also explained by
a similar mechanism. The features we are studying are quite
different, as will be discussed in Section~\ref{sec:1D} (item~9). 

%

\section{WAKE IDENTIFICATION AND REMOVAL}\label{sec:fix}

The wake signature is interesting in itself, and also for what it
can reveal about the solar wind plasma. However, it also constitutes
a major contamination of the measurement of the naturally occurring
electric field. For example, any spectrum of the electric field
will show strong spin harmonics because of the wake. It is therefore
of interest to produce cleaned electric field data, particularly for
the Cluster Active Archive \cite{Lindqvist2005a}, which is 
becoming a major repository of
validated data from all Cluster instruments and 
the major source of Cluster data for the scientific community.

The wake signature depends on the solar wind flow properties
and is therefore usually slowly changing on the Cluster spin
timescale of 4~s. Exceptions to this of course exist,
particularly at boundaries, but for the bulk of the solar wind
data the proposition holds true. This enables us to construct a
"find and fix" algorithm based on determining the wake for a
given spin as the average of the wakes for this and adjacent 
spins. The algorithm have to take care of a lot of special
cases related to data gaps, missing data, interference from
the WHISPER sounder and so on, but its general features are
as follows (see also Figure~\ref{fig:fix}):

\begin{figure}[ht]
\vskip 0.8 cm
\centerline{
	\includegraphics[width=16.0cm]{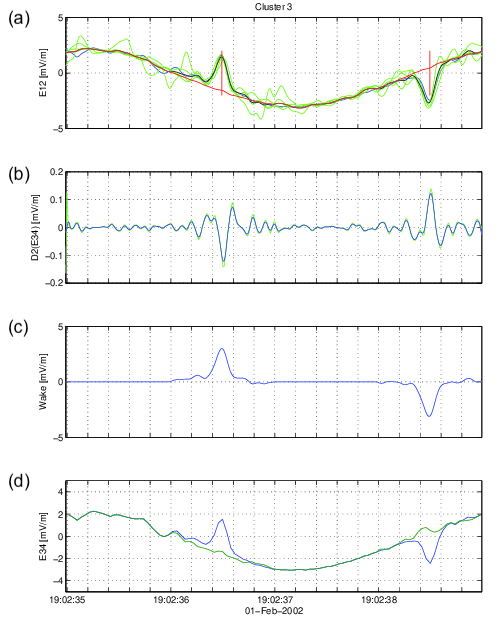}
}
\caption{An example of wake identification and removal
from one spin of EFW data from Cluster~3. (a)~The electric 
field data for the spin under study are plotted in black, with the four
adjacent spins given in green. (b)~The second derivative 
of the weighted average of the five spin periods of
data above. 
(c)~After smoothing of the second derivative, it is integrated
twice to give a wake time series. This is set to zero far from
the wake centre. The amplitude and width of the wake are
determined from these data.
(d)~The original time seried (blue) and the corrected data (green).}
\label{fig:fix}
\end{figure}

\begin{enumerate}
\item
Organize the data per spin.
\item
Resample the data from the original time series (at 25 or 450~samples/s)
to get uniform $1^\circ$ resolution. 
\item
For each spin, form a weighted average of the data from that spin
and its four closest neighbours. We use weights $[0.1, 0.25, 0.3,
0.25, 0.1]$. The wake signature, which is very repetitive
(Figure~\ref{fig:ex}), will not be much changed by this, while
unrelated time variations from wave activity will be damped. 
\item
Calculate the second derivative of the averaged data. This will
exaggerate the wake signature at the expense of the background
sinusoidal spin signature of the natural electric field in the
plasma.
\item
Apply a smoothing filter (we use 7-point moving average) to
minimize the effect of wave noise, which has been boosted by the
differentiation.
\item
Find candidates for wake centre positions 
by looking for maxima of the magnitude of
the second derivative.
\item
Integrate twice to recover a smoothed version of the original
time series.
\item
Take the average of the two wakes found in the spin
(account for the sign) to get an average wake form
for this spin.
\item 
Set the data far from the candidate wake centre positions 
in this time series to zero, so
that only data close to the wake will be affected by the eventual wake
removal. 
\item
Apply restrictions on allowed angle with respect to the solar
direction, amplitude, shape and width of the signature that it
must meet to qualify as a wake.
\item
Remove the wake signature from the original time series of the
spin under study. 
\item
Consider this to be a preliminary wake estimation and cleaning,
that may be
influenced by the fundamental natural electric field signal at
the spin frequency, particularly if the wake occurs close to
an extremum in the spin frequency electric field. To minimize
such errors, we fit a sinusoidal signal at the spin frequency
to the preliminary cleaned data, remove this from the original
data, and perform steps 1--11 again on data that thus have most
of the original spin electric field removed.
\item
Thus we arrive at a final wake estimation which we subtract from
the original data to create a final cleaned
electric field signal. Save characteristic data for the wake:
spin phase angle $\phi$, full width at half maximum (FWHM) $w$, 
and maximum amplitude $A$.
\item
Repeat procedure for next spin.
\end{enumerate}

\begin{figure}[t]
\centerline{
	\includegraphics[angle=-90,width=11.0cm]{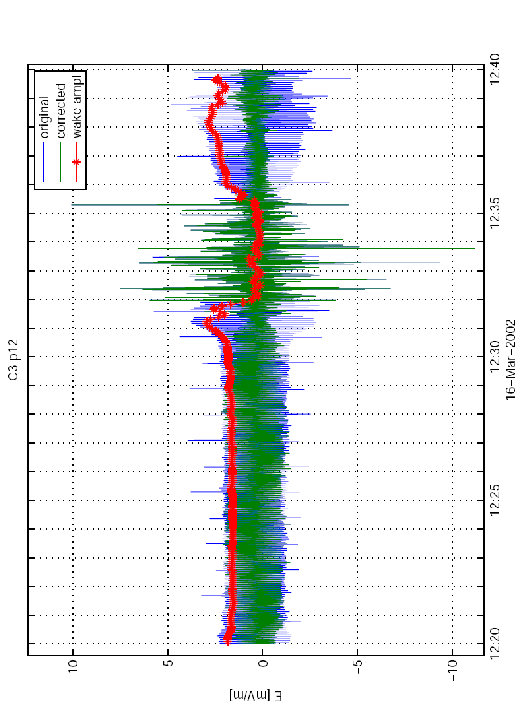}
}
\centerline{
	\includegraphics[angle=-90,width=11.0cm]{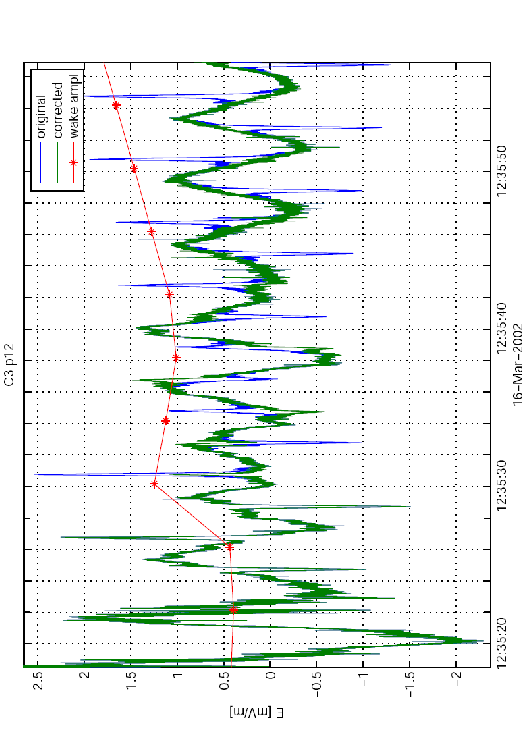}
}
\caption{Full resolution raw (blue) and wake corrected (green) 
  electric field data from EFW probes 1 and 2
  on Cluster~3. The red curve denotes the wake amplitude.
 The lower panel is a zoom in to a detail of the upper panel.}
\label{fig:20mins}
\end{figure}


That this algorithm is quite effective is illustrated by the data
in Figure~\ref{fig:20mins}. The upper panel
shows 20~minutes of full resolution EFW electric field data, 
in the spinning spacecraft
frame, acquired in the solar wind and bow shock regions. The wake
amplitude is very stable for more than 10~minutes, and varies
reasonaby slow even rather close to the activity associated with
the bow shock, which we see from 12:32 to 12:36. A detail of these
data is shown in the lower panel. It can be seeen that the
wake signature appears in the data when exiting the turbulent
region around 12:35:22 -- it may indeed be present inside that
region as well, but is hard to discern even for the human eye
in this dynamic region. The wake is easily spotted from 12:35:24
onward, but it clearly takes a few spins before it stabilizes
sufficiently for the wake algorithm to be able to correctly 
identify the wake pulse in the data and correct for it. However,
it partially alleviates the wake, and hence improve the quality
of any spectrum calculated from these data, even though the
fix is not perfect. But already from 12:35:40 the fix is very good,
and at the end (and for the following hours) it appears almost
perfect to the human eye.

This perfection is of course an illusion. Removing the wake, by
however clever an algorithm, will always be a manipulation of the
data. Any real plasma signals on the same time scale will also
be removed in the process, and natural features from surrounding
spins may be introduced by the averaging. Nevertheless, it is
clear that removing the wake vastly increases the usefulness of
the data, as the major artificial signal (which sometimes can
dominate the time series) is removed.

\section{WAKE STATISTICS}\label{sec:stat}

The algorithm described in Section~\ref{sec:fix} collects data 
on some of three primary charcteristics of the wake: its position
in the spin plane $\phi$, its amplitude $A$ and its width, quantified 
as the FWHM value, $w$. If the shape of the wake
was Gaussian, these parameters would be sufficient to fully describe
it. As the algorithm is used in the production of EFW data for the
Cluster Active Archive, statistics pile up quickly. In the present
study, we limit ourselves to data from Cluster spacecraft 3, and
restrict ourselves to the time period February-April 2003. During
this period, 1,170,617 wakes were identified and corrected for on
Cluster~3, forming a decent statistical sample.

\begin{figure}[t]
\centerline{
	\includegraphics[width=16.0cm]{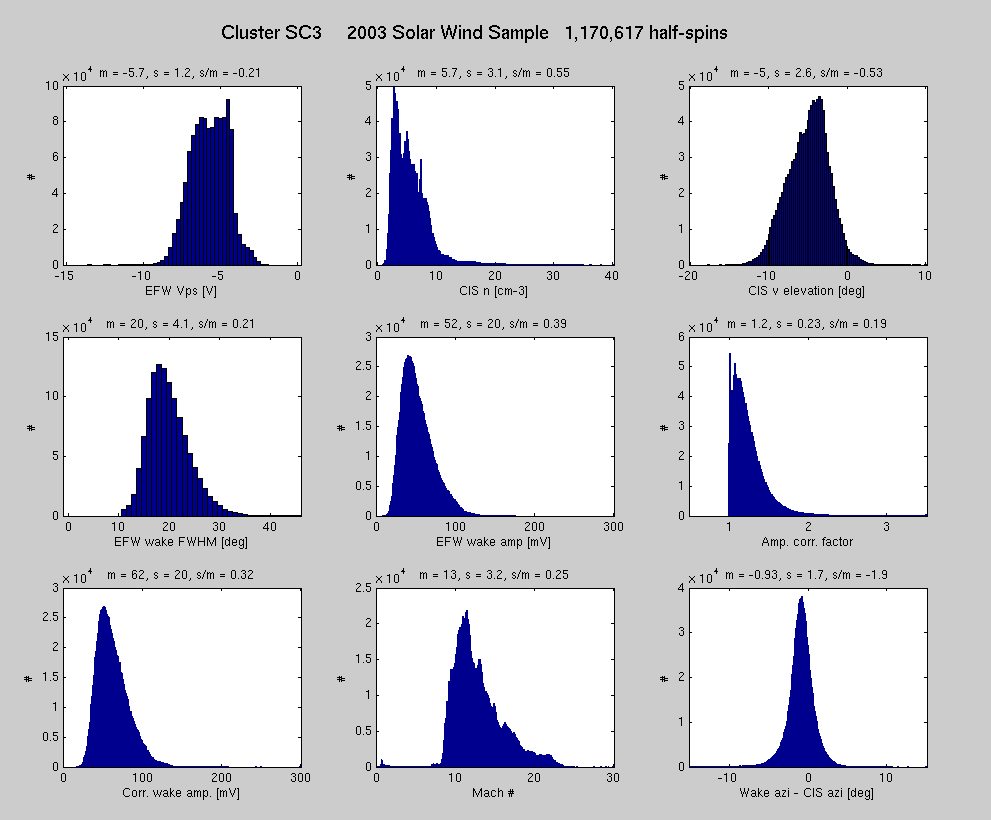}
}
\caption{Statistical distributions of wake parameters and related
  data for the statistical sample used in this paper. On top of each
  panel is given the mean $m$, standard deviation $s$ and normalized
  spread $s/m$ for the plotted variable.}
\label{fig:hist}
\end{figure}

\subsection{1D distributions}\label{sec:1D}

Figure~\ref{fig:hist} shows the observed distributions of the
wake parameters and some complimentary data. The CIS data used
in this study are prime parameters from the Cluster Science Data
System, CSDS \cite{Daly2002a}. We comment the histograms
one by one:
\begin{enumerate}
\item
$V\ind{ps}$, the probe-to-spacecraft potential, can be viewed
as a quasi-logarithmical density proxy 
\cite{Pedersen1995a,Thiebault2006a}, but should also be important
for the wake because it to a large extent determines the 
electrostatic environment around the spacecraft. The distribution
seen is typical for the solar wind \cite{Lindqvist1983a}.
\item
CIS $n$, the plasma density from the CIS HIA detector, shows a
distribution typical for solar wind values.
\item
CIS $\vec{v}$ elevation is the angle of the flow to the spin
plane as observed by the CIS HIA detector. Because the spin
axis is inclined at typically 5--$6^\circ$ with respect to the 
ecliptic plane, the peak is offset from zero, and has a mean
of $5.0^\circ$. The distribution
shows a clear skewness: we have not investigated the reason for
this, or if this is typical for solar wind conditions at other
times.
\item
The EFW wake FWHM value, $w$, has a mean value of $20^\circ$. One
should note that this estimate may be biased because we have 
discarded all wakes with width below $10^\circ$, as such cannot
be reliably determined with the predominating 25~Hz sampling rate
of EFW. Nevertheless, preliminary checks of data acquired at
450~samples/s do not show any big difference from this distribution,
so the number of wakes discarded because of low width is small:
the drop below $18^\circ$ appears to be real.
\item
The EFW wake amplitude, $A$, shows a mean of 52~mV, and a smooth
distribution. Very weak wakes may not be detected, and hence the
lower part of this distribution could be uncertain.
\item
The amplitude correction factor is discussed in Section~\ref{sec:corr}.
\item
The corrected wake amplitude is discussed in Section~\ref{sec:corr}.
\item
The Mach number is here defined as the ratio of the CIS ion flow
speed to the thermal ion velocity, defined as 
$\sqrt{KT\ind{i\perp}/m\ind{p}}$, where $KT\ind{i\perp}$ is the
perpendicular temperature moment from CIS HIA and $m\ind{p}$ is
the proton mass. 
\item
The difference between the wake direction in the spin plane
seen by EFW and the solar wind flow direction shows a very
narrow distribution. The two measurements agree to $-0.9^\circ$
with a standard deviation of $1.7^\circ$. While the mean value
will be discussed in Section~\ref{sec:2D}, we may note that
the standard deviation is as small as it could ever be,
as the typical EFW sampling rate of 25~samples/s corresponds
to an angular resolution of $3.6^\circ$. This impressive
agreement clearly shows that the phenomenon at hand is a
velocity wake, not a magnetic effect as those studied on
Viking \cite{Hilgers1992a} and Akebono \cite{Amano1998a}
(see Section~\ref{sec:ex}). In addition, a plot of the relation between
magnetic field direction and wake direction (not shown)
shows no correlation at all.
\end{enumerate}
 
\subsection{Gaussian wake modelling}\label{sec:corr}

One way to test our understanding of the wake data is to 
investigate the relation between wake amplitude, width and
the flow elevation angle. As a first model, we may assume
that the shape of the wake is Gaussian. This will be the
case if the plasma is Maxwellian, if the electric
potential in the wake
is so small as to not really influence particle motion and
if we are sufficiently far from the spacecraft 
\cite[p.~25]{Alpert1965a}. All these requirements are
probably violated for Cluster in the solar wind, with
the approximation of a "neutral wake" (no influence of
wake electric fields on its structure) being the most
problematic. This can also be seen in the wake signature
in the bottom panel in Figure~\ref{fig:fix}, which suggests
that the wake deviates from a Gaussian shape at least by
having a broad sheath around it. Such a structure is actually
expected, with a sheath characterized by self-similar expansion
surrounding a central wake core where the neutral approximation
applies \cite{Tribble1989a,Murphy1989a}.

Nevertheless, a Gaussian form is a good starting point not
only for this reason, but also
because its factorization properties.
Assume the wake potential is of the form
\begin{equation}\label{gauss}
\Phi(x,y,z) = A_0 f(z)\, \exp \left( -\frac{x^2 + y^2}{a^2} \right)
\end{equation}
where $z$ is the coordinate along the flow direction, the
origin is at the centre of the spacecraft and $a$ is related
to the FWHM by 
\begin{equation}\label{a}
w = 2 a \sqrt{\ln 2}.
\end{equation}
If a probe
cuts the wake through its diameter, as it would do if the
flow was exactly in the spin plane, we put $y = 0$ and the observed 
extremum amplitude is clearly the wake amplitude at that distance, because
$\Phi(x) = A_0 f(z)\, \exp(-x^2/a^2)$. If, however, the flow is out of the
spin plane so that the probe cuts the wake along a chord whose
closest distance to the wake axis is $b$, then 
\begin{equation}\label{chord}
\Phi(x) = A_0 f(z)\, \exp(-b^2/a^2)\, \exp(-x^2/a^2).
\end{equation}
Note that the observed FWHM value is independent of $b$, as we
derive it from a series of data with varying $x$. Hence, if we know
that the flow direction is elevated by an angle $b$ from
the spin plane, we expect the real amplitude at distance $z$
from the spacecraft, $A_0 f(z)$, to be related to the amplitude 
we observe, $A = \max(\Phi(x))$, by
\begin{equation}\label{corr}
A_0 f(z) = A\, \exp(b^2/a^2) 
\end{equation}
if the wake is assumed to be Gaussian in the angles. This is
in fact a substantial differenece from being Gaussian in
the Cartesian coordinates as we saw in Section~\ref{sec:1D}
that the wake width (FWHM) shows a mean as large as $20^\circ$,
and can reach above $30^\circ$ at times. Nevertheless,
as the EFW probes are at constant distance $r = 44$~m from
the spacecraft, not at constant downstream coordinate $z$,
assuming Gaussian shape in the Cartesian coordinates would
mean that we had to enter some assumption on $f(z)$, so we
here stick to the angular Gaussian as a first approximation.
It therefore makes sense to correct the wake potential for
the inevitable underestimate due to the probe not passing
through the centre of the wake by means of Equation~(\ref{corr}),
with $a$ calculated from the observed FWHM as in Equation~(\ref{a})
and $b$ is the CIS flow elevation angle.

The distribution of this amplitude correction factor, 
$\exp(b^2/a^2)$, is plotted
in the right panel in the centre row of Figure~\ref{fig:hist}.
The distribution of the corrected amplitudes, $A\, \exp(b^2/a^2)$,
is shown in the left panel of the lower row. Comparing to
the uncorrected amplitude in the centre of the middle row,
we find that the spread, defined as the ratio of the
standard deviation $s$ to the mean $m$, has decreased from
0.39 to 0.32. That the spread decreases shows that our
basic understanding of the wake geometry is correct: had the
amplitude, width and flow elevation not been connected by
a relation like (\ref{corr}), the spread would not have
decreased by the application of this correction factor. While
this does not imply that the wake is exactly Gaussian, it demonstrates
that the cause of the pulse-like signatures is a wake with a
similar geometric structure.

Note that our assumption of a Gaussian wake is only used for
deriving a corrected wake amplitude: the rest of this study
is independent of this assumption.

\begin{figure}[t]
\centerline{
	\includegraphics[angle=-90,width=17.0cm]{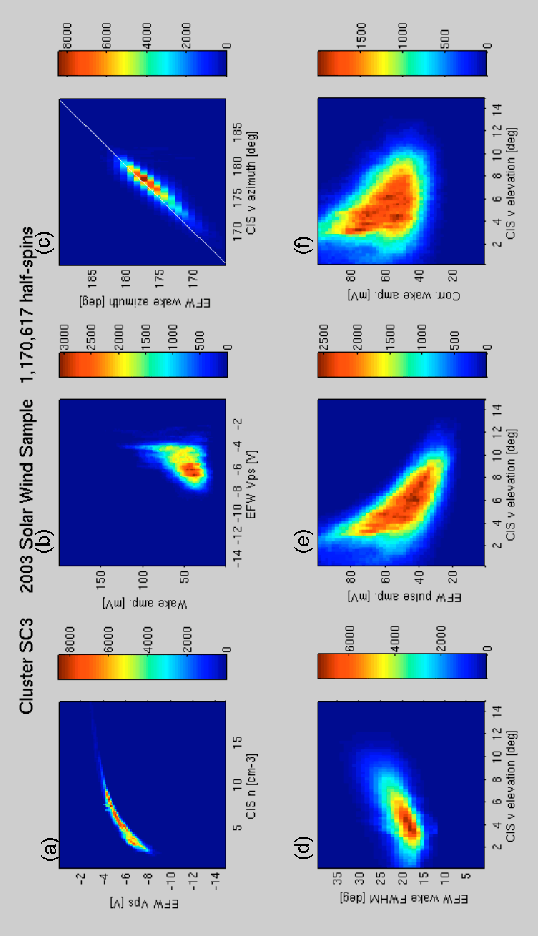}
}
\caption{Two-dimensional histograms of wake and plasma parameters.
Colour scales give the number of samples in each bin.}
\label{fig:2D}
\end{figure}

\subsection{2D distributions}\label{sec:2D}

To further see the correlation between the wake parameters,
we present two-dimensional histograms in Figure~\ref{fig:2D}.
Panel~(a) illustrates how the probe-to-spacecraft potential $V\ind{ps}$,
which is linearly related to the spacecraft potential, depends
on the plasma density as measured by CIS. This relation is
further discussed elsewhere in this volume \cite{Eriksson2007c},
and is often invoked for using $V\ind{ps}$ as a density proxy
\cite{Pedersen1995a}.

Panel~(b) shows the variation of wake potential amplitude with 
$V\ind{ps}$, and hence with density. An increasing
trend can be discerned, which can be understood as an
effect of the Debye length for the charging of the wake.
If we remove all ions but keep all electrons in a cylindrical
region of a plasma, the resulting potential will scale as
\begin{equation}
\Phi \sim \frac{K T\ind{e}}{e} \left( \frac{R}{\lambda_D} \right)^2
\end{equation}
for wake radii $R \lesssim \lambda_D$. Typical Debye lengths
in the solar wind are 5--10~m, and the transverse size of the
depleted region should not be too different from the spacecraft
dimensions even as far downstream as the EFW bbom length of 44~m,
so we should indeed be in this regime. Shorter Debye lengths
should thus give higher wake potentials, as observed.

Panel~(c) is a deeper look at the relation betwwen the wake
direction from EFW and the flow direction from CIS. While the
points obviously line up nicely, there is a small but very
consistent deviation from the straight line of slope one as we go
away from the purely antisunward direction ($180^\circ$). This
means that the wake direction from EFW data is a few degrees
more sunward than the flow direction from
CIS. It is clear that this is the reason for the small negative
mean we saw in Section~\ref{sec:1D}, Item~9.
We do not know the origin of this phenomenon. It may
be due to the photoelectron cloud around the spacecraft,
which is primarily controlled by the direction to the sun rather
than the flow direction, influencing the wake structure, or be
due to some artefact arising from the EFW or CIS measurements.

In Panel~(d) of Figure~\ref{fig:2D}, we find a clear deviation
from what we would expect for a perfectly Gaussian wake 
(Section~\ref{sec:corr}). While Equation~(\ref{chord}) implies
that the observed width $w$ should be independent of flow
elevation angle over the spin plane, Panel~(d) shows that
this is not the case.  

Finally, Panels~(e) and (f) in Figure~\ref{fig:2D} show the pulse 
amplitude detected by EFW. Panel~(e) shows the general
characteristic expected by an axially symmetric wake
aligned with the flow direction: the probes on the wire boom
on the spinning satellite cut a chord throguh the wake, and
the further the chord is from the centre axis, the smaller
is the pulse amplitude detected. In Panel~(f), we show the
result of applying the amplitude correction factor defined
by Equation~(\ref{corr}), which assumes a Gaussian wake structure.
Comparing to Panel~(e), we see a partial but certainly not
complete removal of the dependence of the amplitude on 
flow elevation angle. This leads to the same conclusion as
in Section~\ref{sec:corr}: the Gaussian assumption catches
some general aspects of the wake, but the detailed shape
of the wake is not well described by a Gaussian.

\section{NUMERICAL SIMULATION WITH SPIS}

The data presented here makes clear the nature of the pulses
in electric field data in the solar wind, and 
the algorithm described in Section~\ref{sec:fix} does a good
job in cleaning EFW data from the influence of the wake.
Hoewever, there are still reasons for further study of the 
solar wind wake. If we have
a good theoretical model of the wake, it may be possible to
use determine plasma parameters like the ion and/or electron
temperatures from the wake measurements, by inversion of the
model or fits to it. Such theoretical understanding will
require better models than the simple Gaussian we used in
Section~\ref{sec:corr}, and would need to be compared to
numerical simulations, or possibly be empirical parametrizations
of results from such simulations.

\begin{figure}[t]
\centerline{
	\includegraphics[angle=-90,width=12cm]{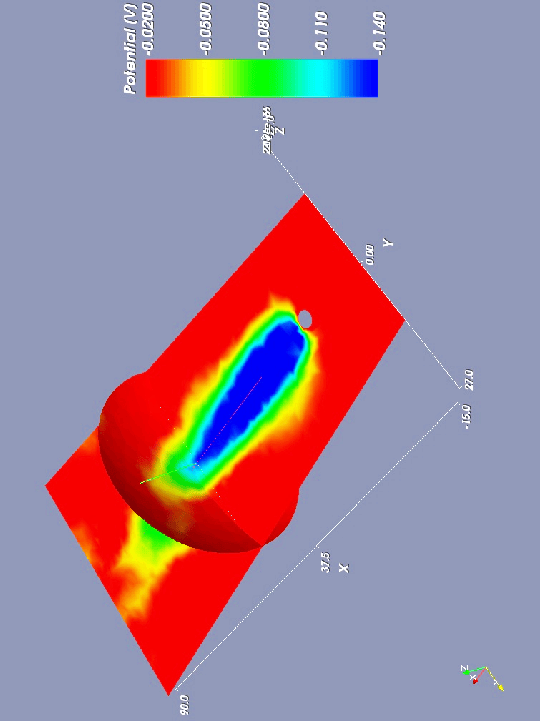}
}
\caption{Wake potential from a SPIS simulation. The solar wind flow
is along the $x$ direction, and the satellite spin axis is along
$z$. The potential is colour coded on the $xy$ plane and on a
spherical surface with radius 44~m. The latter represents the
wake region crossed by th EFW probes, at different values of $z$
depending on the flow elevation angle over the spin plane.}
\label{fig:spis}
\end{figure}

In Figure~\ref{fig:spis}, we show a result from a first numerical
simulation using the SPIS simulation package, available at
{\tt http://www.spis.org} and described in several papers
\cite{Roussel2005a,Forest2005a,Hilgers2005a}. SPIS is a
comprehensive package running on a multitude of platforms,
which makes it very useful for desktop simulations like this, 
where only a small 
amount of manpower can be invested in the simulation.
The simulation presented here was run on an ordinary 1.8~GHz Pentium~M
linux laptop computer with 1.5~GB of RAM, on which it completed
within 20~hours.

The simulated plasma is typical for the
solar wind: number density 10~cm$^{-3}$, electron temperature
10~eV, ion temperature 5~eV, and flow speed 310~km/s. The ions
are all assumed to be protons. In such a plasma, the Cluster
spacecrafts would usually attain a spacecraft potential of
4--5~V, as can be seen fron Figure~\ref{fig:2D}a. However, in
this simulation we have put the spacecraft potential to zero.
This makes it much simpler to extract the potential due to the
space charge in the wake, as we do not have to subtract any vacuum 
field, and should not introduce any appreciable errors to the
wake structure: the protons all arrive at energies around 500~eV,
and would thus care little about a few volts of potential anyway.
The electron distributions will be affected, but only very close 
to the spacecraft, so that any impact on the distant wake from
this assumption will be negligible.
We have set SPIS to use a full PIC implementation for
the ions, while the electrons are assumed to be Boltzmann
distributed.

The simulation region is a circular cone with its top chopped
off, so it has circular cross sections at the
downstream and upstream ends. The upstream circle has a radius
of 13~m and is 15~m in 
front of the spacecraft, while the downstream boundary is 27~m
in radius and is located 90~m downstream of the satellite body.
The spacecraft is a circular cylinder, 1.5~m in radius and
1~m in height, centred at the origin with its axis along the $z$ axis,
while the flow is in the positive $x$ direction. The 
characteristic dimensions for the automatic gridding are set to 
3~m on the outer boundaries and 0.5~m on the spacecraft.
This resulted in a division of the simulation volume into
just under 46,000 tetrahedra.

The Poisson equation is solved
using an implicit scheme, and some under-relaxation is used
to stabilize the plasma calculations. The total simulation time
is 1~ms, selected to be roughly three times the solar wind propagation time
through the simulation region. The average number of
macroparticles per cell is set to 60. The number of macroparticles
was 1.85~million at the start of the simulation and 2.63~million at
the end. No photoelectrons or
magnetic fields are included in the simulation: this is justifiable
as most photoelectrons will be retained close to the spacecraft
by its positive potential, and the relevant particle gyroradii 
in typical solar wind magnetic fields (some 5~nT)
are larger than the simulation volume. 

\begin{figure}[t]
\centerline{
	\includegraphics[width=10cm]{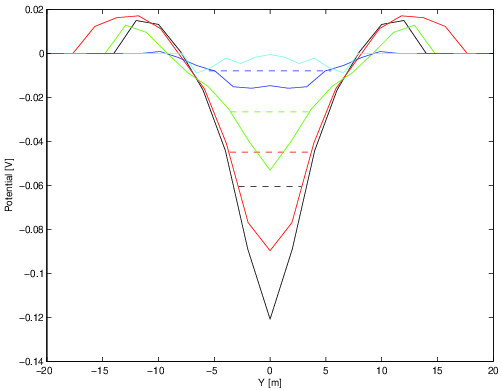}
}
\caption{Wake potential profiles 44~m downstream of the spacecraft  
from the SPIS simulation, obtained for
different solar wind flow elevation angles above the spin plane:
$0^\circ$ (black), $5^\circ$ (red), $10^\circ$ (green), 
$15^\circ$ (blue) and $20^\circ$ (cyan). The half-maximum
levels are indicated by dashed horizonta lines.
}
\label{fig:cut}
\end{figure}

Figure~\ref{fig:spis} clearly shows 
a wake forming behind the satellite, charging
to about -140~mV at the spherical shell at 44~m distance where the
EFW probes will make their measurements. For zero flow elevation,
each probe crosses the wake along the intersection of the planar
and spherical surface elements in Figure~\ref{fig:spis}. For nonzero flow
elevations, the probe moves along the intersection of the spherical
surface with a plane through the spacecraft, similar to the plane
shown but turned around the $y$ axis by the elevation angle. This 
wake cut may be approximated by a cut of the spherical
segment at constant nonzero $z$. We present such cuts in
Figure~\ref{fig:cut}. While some general similarity with the
example observation in Figure~\ref{fig:ex}c is clear, 
a more detailed investigation
of the shapes of the observed wakes as well as a larger number
of simulations with refined grid resolution would be necessary
for a detailed comparison. One interesting
feature is the local maxima on the edges of the wake seen 
in the simulation. There is no obvious such structure in the
example data in Figure~\ref{fig:ex}c: it would be interesting
to study a larger number of wake shapes to see if they
turn up only in some cases, or never at all. The simulated 
wake amplitudes at 44~m are around 90~mV for $5^\circ$ flow
elevation and 50~mV for $10^\circ$. These values lay in the
upper range of what we would expect from Figure~\ref{fig:2D}e,
but we should note that a careful comparison also will need
to take into account the effect of the 10~Hz anti-aliasing
filters and 25~Hz sampling frequency of the EFW instrument,
which should be applied to the simulation data in order
to make them fully comparable.

\section{CONCLUSIONS}

In this paper, we have described the formation of wakes behind
spacecrafts in the solar wind. We have shown
the signature of the wake in data from electric field 
measurements on the Cluster satellites,
and how to separate the contribution from the wake from
the natural electric field. We presented statistics of the
wake based on a large sample of over 1~million wake observations,
showing general consistency between expected and observed wake
characteristics. We have verified the wake formation in a 
numerical simulation with SPIS and discussed its possible use 
for determining plasma parameters.  
Our principal results can be summarized as follows:

\begin{enumerate}

\item
The Cluster electric field instrument EFW detects pulse-like
signatures in the solar wind, repeated at spin frequency.
\item
These structures are shown to be caused by
a wake behind the spacecraft: they align with the flow,
and their amplitude decreases as the flow elevation angle
above the spin plane increases.
\item
We have constructed an algorithm for detecting and removing
the wakes from the electric field data.
\item
The wake formation is confirmed by SPIS simulations.
\end{enumerate}

\section*{Acknowledgements}
We thank the CIS team (PI Henri R\`{e}me) for the HIA particle data. 
We also thank the SPIS developers, the SPIS sponsor ESA and the 
SPINE network for creating not only the SPIS code but also a 
stimulating scientific and human environment around it.
This study was supported by the Swedish Space Board
(Cluster EFW operations and data analysis) and
by ESA (the EFW part of the Cluster Active Archive).

\bibliographystyle{unsrt}
\bibliography{ref}

\end{document}